\documentclass[12pt,twoside]{article}
\usepackage{xrb2007}
\usepackage[dvips]{graphicx,rotating}
\DeclareGraphicsExtensions{.ps}

\newcommand{\rxte}{\textit{RXTE}}
\newcommand{\inte}{\textit{INTEGRAL}}
\newcommand{\igr}{IGR~J19140$+$0951}

\begin{document}

% select your session by uncommenting the appropriate line
\session{Obscured XRBs and INTEGRAL Sources}

\shortauthor{Prat, Rodriguez and Hannikainen}
\shorttitle{Observations of \igr}

\title{\rxte\ and \inte\ observations of \igr}
\author{L. Prat, J. Rodriguez}\affil{Laboratoire AIM, CEA/DSM - CNRS - Universit\'e Paris Diderot, DAPNIA/Service d'Astrophysique,Bat. 709, CEA-Saclay, F-91191 Gif-sur-Yvette Cedex, France}
\author{D.C. Hannikainen}\affil{Observatory, PO Box 14, FI-00014 University of Helsinki, Finland}

\begin{abstract}
We have used the \rxte\ and \inte\ satellites simultaneously to observe the High Mass X-ray binary \igr. The spectra obtained in the 3--80 keV range have allowed us to perform a precise spectral analysis of the system along its binary orbit. The spectral evolution confirms the supergiant nature of the companion star and the neutron star nature of the compact object. Using a simple stellar wind model to describe the evolution of the photoelectric absorption, we were able to restrict the orbital inclination angle in the range 37--75 degrees. This analysis leads to a wind mass-loss rate from the companion star of $\sim$ 10$^{-7}$ M$_{\sun}$/year, consistent with its expected spectral type. We have detected a soft excess in at least three observations, for the first time for this source. Such soft excesses have been reported in several HMXBs in the past. We discuss the possible origin of this excess, and suggest, based on its spectral properties and occurrences prior to the superior conjunction, that it may be explained as the reprocessing of the X-ray emission originating from the neutron star by the surrounding ionised gas.
\end{abstract}

\section{Introduction}
High Mass X-ray Binaries (HMXBs) are binary systems consisting of a compact object orbiting a massive companion star. Prior to the launch of the INTErnational Gamma-ray Astrophysics Laboratory (\inte) in 2002, most of the known HMXBs contained a Be companion. In Be-type HMXBs, the compact object emits strong X-ray flashes when it crosses the equatorial plane of the companion star, where a thick disk of matter originating from the stellar wind is present. O and B-type stars have more isotropic stellar winds which absorb the X-ray emission of the compact object, rendering them almost undetectable below a few keV. Since the launch of \inte, however, many such systems have been found \citep[see e.g.][]{IGR_1725}, raising the ratio of Be-HMXBs to O,B-HMXBs close to 0.5. The study of this new population may have strong implications on the scenarios of binary formation and evolution. In this perspective, the use of X-ray spectroscopy at different orbital phases makes it possible to probe the stellar wind, providing two-dimensional information on the density and ionization structure of the wind. For instance, the soft excess that is present in the soft X-ray spectra of many HMXBs, whose origin is still quite mysterious, is linked to the physics of the wind close to the compact object, especially the region where the fast moving stellar wind collides with the slow moving and highly ionized gas surrounding the compact object \citep{soft}.

\igr\ was discovered by \inte\ by the satellite \inte\ in March 2003 \citep{decouverte}. Prior to our study, it was identified an a HMXB with an orbital period of 13.552 days \citep{periode}, composed probably of a neutron star \citep{Jerome} orbiting a B0.5 I type supergiant companion star \citep{Hanni}. Thus, it is a good candidate to study the properties of the obscured X-ray binaries.

\section{Soft Excess detection}

The data from the two satellites were reduced using the latest software packages available. Since \igr\ is a fairly faint source, we also corrected the spectra for the Galactic X-ray background (GXB), using the parameters measured by \citet{Valinia}. This correction aimed mostly at minimising the errors on the measure of the photoelectric absorption, by removing the Galactic contribution and keeping only the absorption intrinsic to the source. We identified the emission process as thermal Comptonization, which is quite common for this type of source. But a more interesting characteristic is the detection of a so called "Soft Excess".

During our study, some spectra exhibited an excess in the soft X-ray part of the spectra, which we modeled by adding a black body component to the model. This feature has already been observed in many X-ray binaries \citep[see][for a review]{soft}. Its origin is mysterious, but can be linked to the immediate environment of the neutron star, and explained qualitatively by the following scenario. Due to its high energy emission, the neutron star ionizes the surrounding material. Since the stellar wind is accelerated by ionization in the UV lines, the already ionized gas around the neutron star is no longer accelerated by the stellar radiation field. When the compact object moves along its orbit, the hot gas will gradually be overtaken by the stellar wind. This will lead to the formation of a ``tail" trailing the neutron star. 

As the gas in this tail is ionized, its absorption will be lower, but its precise effect on the emission is not clear. Since the Soft Excess is only visible just prior the superior conjunction (between phases 0.25 and 0.5), one possibility is that the hard X-ray emission may be scattered around the tail. The precise study of this effect, however, would require better spectral resolution in the soft X-ray range than what is available on \rxte.

\section{Wind model}

\begin{figure}
	\begin{center}
	\includegraphics[width=8cm]{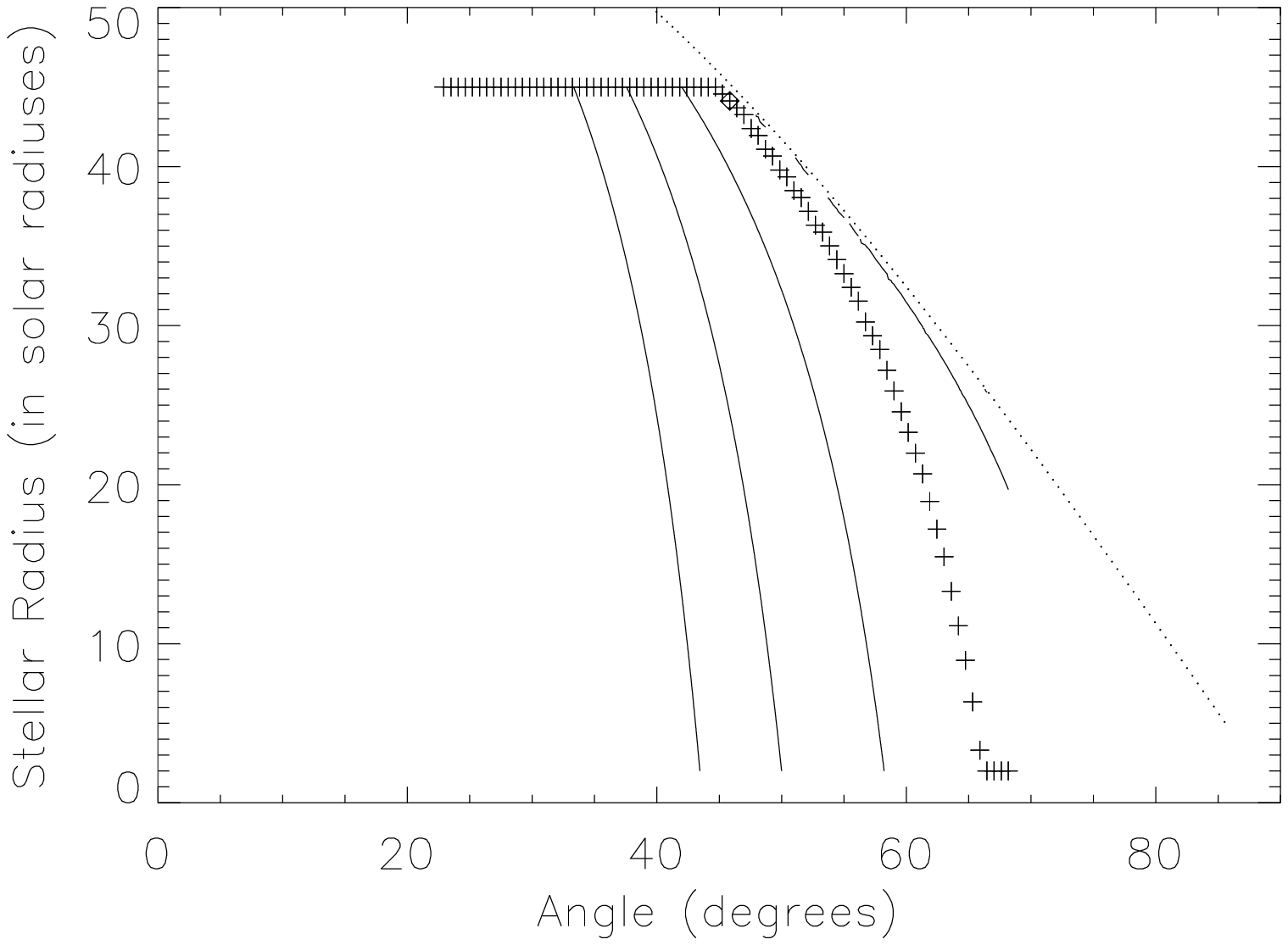}
	\includegraphics[width=7.5cm]{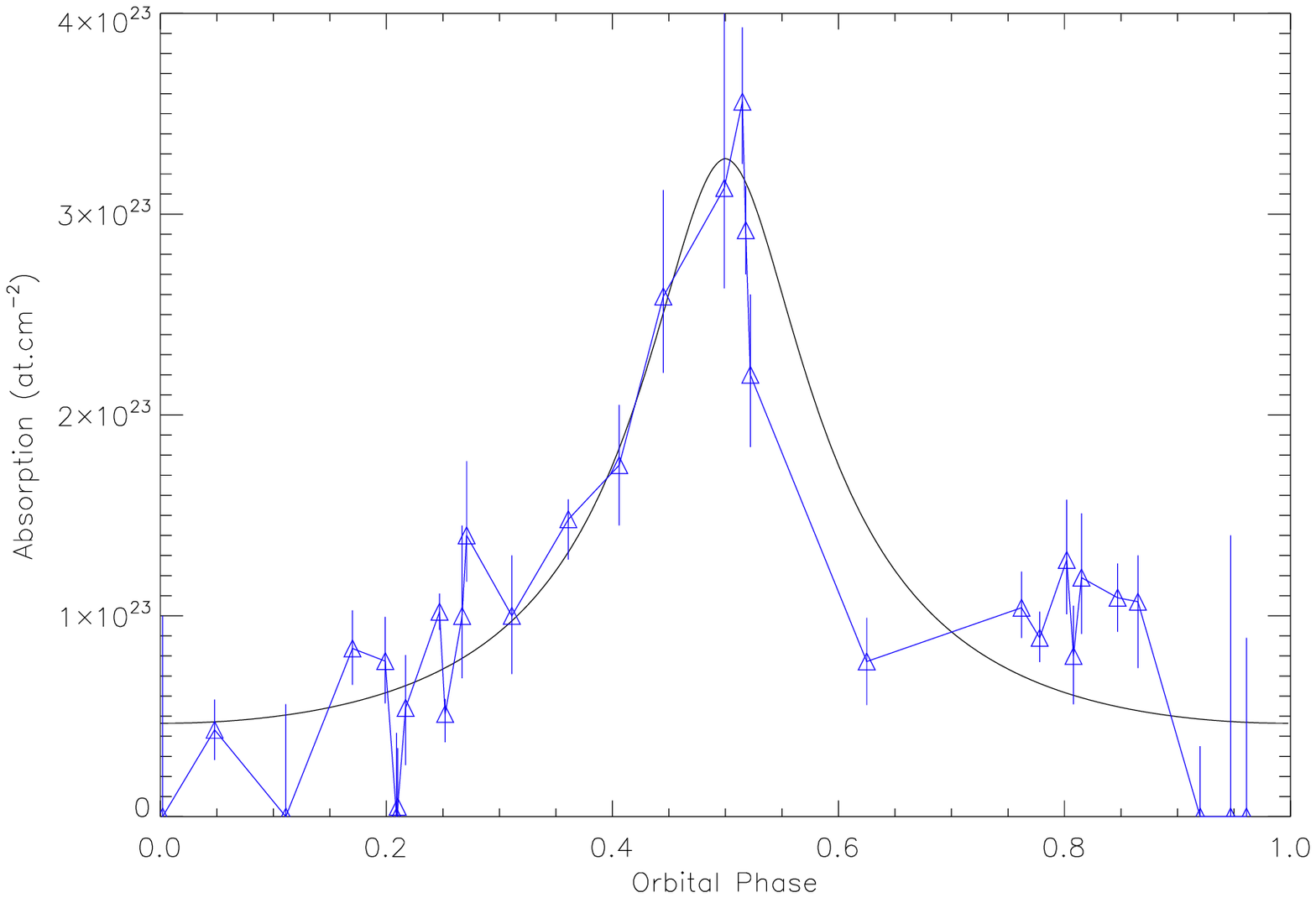}
	\caption{{\bf{Left :}} Most probable orbital inclination, as a function of the companion radius, for a typical set of parameters. The + symbols correspond to the most probable orbital inclination, confidence contours are drawn at the 25, 68 and 90\% confidence levels. The dotted line is the ``eclipse limit": points above this line are excluded as it would imply an eclipse of the neutron star by its companion, which is not observed. {\bf{Right :}} Model absorption (continuous line) and experimental absorption ($\triangle$ symbols), for the stellar model with parameters M$_{\star}$ = 20M$_{\sun}$ , R$_{\star}$ = 21R$_{\sun}$ and $\beta$ = 0.5. Phase 0 corresponds to when the compact object is located between the Earth and the companion star (inferior conjunction).}
	\label{meilleur}
	\end{center}
\end{figure}

In order to constrain several parameters for the system, we used a simple wind model based of the fact that B-type stars emit strong stellar winds, usually taken to be stationary and spherically symmetric. See e.g. \citet{inclinaison} for a precise description of the method. Using the evolution of the photoelectric absorption, we were able to constrain the inclination of the system and the mass-loss rate of the companion star. For the most probable stellar radius of 21 $R_{\sun}$, the lower inclination limit is constrained between 37 and 42 degrees (fig. \ref{meilleur}, left), with $\chi^2$ being at a minimum in the range 55--63 degrees. Fig. \ref{meilleur}, right, shows the best-fit model against the experimental normalised data, with a good agreement. $\dot{M}_{\star}$ is constraint in the range 0.6--1.0$\times$10$^{-7}$ M$_{\sun}$/year, which is consistent with what is expected for a B0.5 type supergiant.

\section{Conclusions}

The study of X-ray binaries is challenging since it is often difficult if not impossible to identify their visible and infrared counterparts. Even if an infrared counterpart were observed, the distance to some systems prohibits the measurement of the orbital characteristics. Our study shows that X-ray observations can overcome these limitations and produce very precise inferences. The \rxte\ and \inte\ observations of \igr\ have led to good measurements of the orbital period of the system and constraints on its inclination angle. 

Moreover, we can use the compact object to probe the stellar wind of the companion. In the case of \igr, we diagnosed the type of the companion (supergiant O or B), the wind density and its structure around the neutron star. More precise observations could lead to constraints on the mass and radius of the companion, and better constraints on the stellar wind. This allows the study of new \inte\ sources, either distant or highly absorbed, and ultimately the determination of new useful data for X-ray binary evolution scenarios.

\end{document}